\documentstyle[aps,twocolumn,psfig]{revtex}

\title{Continuously induced coherence without induced emission}

\author{J. \v{R}eh\'{a}\v{c}ek$^{\ast}$, L. Mi\v{s}ta Jr.$^{\ast}$,
J. Fiur\'{a}\v{s}ek$^{\ast}$ and
J. Pe\v{r}ina$^{\ast\dagger}$\\ \it
$^{\ast}$ Department of Optics, Palack\'{y} University, 17. listopadu 50,
772 00, Olomouc, Czech Republic\\
$^{\dagger}$ \it Joint Laboratory of Optics of Palack\'{y} University and
Physical Institute of Czech Academy of Sciences,\\
\it 17. listopadu 50, 772 00, Olomouc, Czech Republic}

\date{}

\begin{document}
\maketitle

\begin{abstract}
The coherence properties of two downconversion processes
{\em continuously} coupled via idler beams are analyzed.
We find that the amount of which-way information
about a signal photon carried by idler beams periodically
attains its maximum and minimum in the course of evolution.
In correspondence with the famous experiments by Mandel's group
on the induced coherence without induced emission the coherence
of signal beams is governed by that information.
The ideal which-way measurement is constructed.
\end{abstract}

\section{Introduction} \label{sec-introd}

In famous experiments by Zou {\em et al.\/} and
Wang {\em et al.\/} \cite{mandel}
it has been shown that the coherence of signal photons
originating in two downconversion processes
can be induced by making them indistinguishable.
This has been accomplished by aligning the idler beam
of the first process with the idler beam of the other,
preventing the possibility to tell the signal photons
originating in different crystals from each other
based on the observation performed on the idler beams.
Aligning the idler beams in this experiment in fact amounts
to introducing a discrete (temporally localized)
coupling of the two processes -- the output
of the first one being the input
of the other. However, a discrete coupling
is not the only possibility here.
It is well known that many interesting dynamical and
quantum-statistical phenomena arise as well from the
{\em continuous}
coupling of two or more optical processes \cite{couplers}.
A~natural question arises whether also the interplay
between quantum indistinguishability and coherence
is reflected in the dynamics of two downconvertors
experiencing continuous rather than discrete coupling.
Such scheme would be a generalization of the original experimental
setup of Zou {\em et al.}\/ \cite{mandel} which might exhibit richer
phenomena. For example, one could imagine that the which-way information
carried by the output idler beams of continuously
coupled downconvertors could somehow depend on the length of
the interaction region.

The analysis will be done in two steps.
First, the coherence properties of two continuously
coupled downconversion processes will be
discussed for a particular choice of the relevant parameters.
Then the setup will be replaced by two simpler devices --
one resembling the experiment of
Zou {\em et al.}\/ \cite{mandel} and the other
being the two-photon interferometer of Ou {\em et al.}\/ \cite{ou} --
performing the same input-output transformations, which
are more convenient for a general analysis in terms of
which-way information.

\section{Induced coherence without induced emission}
\label{sec-induced}

\begin{figure}
\centerline{\psfig{file=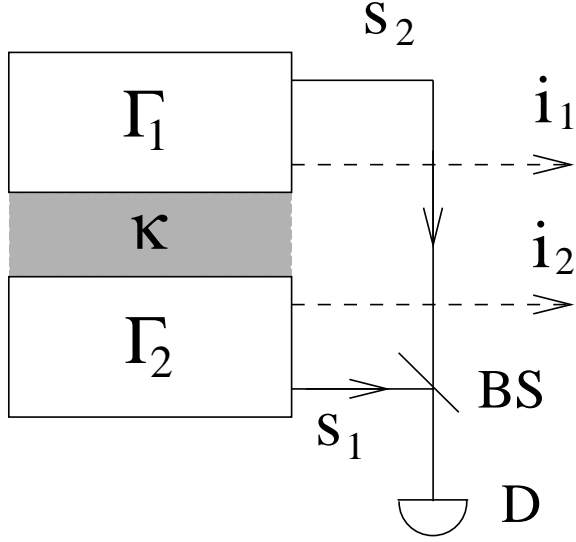,width=0.99\linewidth}}
\caption{Experimental arrangement; gray color
indicates the linear interaction region; BS -- beamsplitter,
D -- detector.}
\label{fig-coupler}
\end{figure}

The setup of the thought experiment which will be discussed in the
following is shown in Fig.~\ref{fig-coupler}.
Two spontaneous downconvertors fed by strong coherent sources
generate four downconverted modes, hereafter denoted
$s_1$, $i_1$, $s_2$, $i_2$; $s$ and $i$ stands for
``signal'' and ``idler'', respectively.
Rather than cascading the downconvertors and aligning
the idler beams as was proposed and realized by Zou {\em et al.}\/
\cite{mandel} we let the idler beams interact
continuously.
The conceptually simplest interaction between two modes
is a linear energy exchange, which can be easily
realized e.g. by means of evanescent waves \cite{coupl}.
At the output, the signal beams are superimposed
at beamsplitter BS and then detected by detector $D$.
The resulting interference pattern is scanned by varying
the signal path difference.

The experimental setup in Fig.~\ref{fig-coupler} (without
beamsplitter BS) can be thought of as a single
device described by the effective interaction momentum operator
\begin{equation}\label{strmom}
\hat{G}=\hbar(\Gamma_1\hat{A}_{s_1}^{\dag}
\hat{A}_{i_1}^{\dag}+\Gamma_2\hat{A}_{s_2}^{\dag}
\hat{A}_{i_2}^{\dag}+\kappa \hat{A}_{i_1}\hat{A}_{i_2}^{\dag}+
{\rm h.c.}).
\end{equation}
Here $\hat{A}_k$ is the annihilation operator of mode $k$,
$\Gamma_1$ and $\Gamma_2$ are the effective strengths of the
corresponding downconversion processes \cite{hong85}
and $\kappa$ is the strength of the linear interaction acting
between modes $i_1$ and $i_2$. The unitary evolution operator
generated by $\hat{G}$ reads $\hat{U}=\exp[i\hat{G}z/\hbar]$
and the spatial evolution of field operators is governed by
the Heisenberg equations of motion
\begin{eqnarray}\label{strequat}
\frac{d\hat{A}_{s_1}}{dz}=i\Gamma_1\hat{A}_{i_1}^{\dag},&\quad&
\frac{d\hat{A}_{i_1}}{dz}=i\Gamma_1\hat{A}_{s_1}^{\dag}+
i\kappa^{\ast}\hat{A}_{i_2},\nonumber \\
\frac{d\hat{A}_{s_2}}{dz}=i\Gamma_2\hat{A}_{i_2}^{\dag},&\quad&
\frac{d\hat{A}_{i_2}}{dz}=i\Gamma_2\hat{A}_{s_2}^{\dag}+
i\kappa^{\ast}\hat{A}_{i_1},
\end{eqnarray}
The device can operate in two very different regimes:
\begin{eqnarray} \label{rezimy}
|\kappa|<|\Gamma_1|+|\Gamma_2|&\quad& \mbox{above threshold},
\nonumber \\
|\kappa|>|\Gamma_1|+|\Gamma_2|&\quad& \mbox{below threshold}.
\end{eqnarray}
Above the threshold the intensity of downconverted light
grows exponentially with increasing interaction length $L$,
whereas below the threshold all modes exhibit oscillations.
We will mainly be interested in the coherence of the signal beams.
We define the (normalized) mutual coherence of signal beams
$\gamma$ through the relation
\begin{equation} \label{coherence}
i\gamma\equiv \frac{\langle \hat{A}^{\dag}_{s_1}
\hat{A}_{s_2}\rangle}{\sqrt{
\langle\hat{A}^{\dag}_{s_1}\hat{A}_{s_1}\rangle
\langle\hat{A}^{\dag}_{s_2}\hat{A}_{s_2}\rangle}},
\end{equation}
where the imaginary unit $i$ has been factored out in order
to make $\gamma$ a real quantity.
In principle there are two ways how the signal beams
originating in two spontaneous downconvertors can
become coherent. Either the coherence can be induced by the emission
of signal photons stimulated by idler photons
traversing from one medium to the other one, or,
as was shown in \cite{mandel}, it can originate in the principal
indistinguishability  of signal photons.
In experiments of Zou {\em et al.}\/ the former (classical)
cause of coherence was eliminated simply by keeping the low rate
of generation of downconverted photon pairs.
This can be easily imitated with our setup in Fig.~\ref{fig-coupler}.
We have already mentioned that below threshold the
intensities of all the four downconverted modes oscillate.
It can be shown that the amplitude of the oscillations
decreases with increasing strength $\kappa$ of the linear
interaction \cite{reh-zeno}. This is caused by
the effective phase mismatch introduced by the continuos
interaction (see also the discussion in section~\ref{zeno}).
We note that such an inhibition of downconversion process
by coupling it to another mode or process
can alternatively be interpreted as the quantum Zeno effect
\cite{others,luis},
the linear coupling being a sort of continuous measurement
\cite{MPS,reh-zeno}. Hence for sufficiently strong
coupling strength $\kappa$, the intensities of all the
involved modes keep low, and spontaneous emission dominates
in the course of evolution.

A~typical behavior of the mutual coherence function
(\ref{coherence}) well below threshold is shown in
Fig.~\ref{fig-coherence} (solid line).
A~characteristic resonant feature can be seen in Fig.~\ref{fig-coherence}
repeating regularly. The coherence of the signal modes periodically
attains its maximum allowed values $\pm 1$ and hence the signal
modes become fully coherent every now and then. We remind the
reader that this coherence is not induced by a stimulated emission
because the rate of the emissions stimulated by the exchanged idler photons
remains very small compared to the rate of spontaneous emissions. This
is due to the small mean number of
photons present in the system (see Fig.~\ref{fig-coherence}, dotted line).
Analogously to Zou {\em et al.}\/ \cite{mandel}
we claim that for the corresponding
interaction lengths the state of the output light is such that
no matter what measurement is performed on the output idler modes,
no information about the mode in which the signal photon has left
the coupled downconvertors is gained; the probability of
having more than one output
signal photon being very small. Similarly it is tempting to
state that roots of $\gamma$ imply the presence
of a perfect which-way information about the signal photon
in the idler modes for these interaction lengths.
This would correspond to preventing the first
idler beam from reaching the second crystal in the experiment
by Zou {\em et al.\/} \cite{mandel}.

\section{Substituting schemes}
\subsection{Modified setup of Zou {\em et al.}\/}

Although it is possible to obtain analytical solutions of
the system (\ref{strequat}), they are awkward
and not suitable for physical discussion.
We will adopt a different strategy.
With the assumption that the strengths $\Gamma_1,\Gamma_2$ and
$\kappa$ are real (without loss of generality), the momentum operator
(\ref{strmom}) can be written as a linear combination
\begin{equation}\label{rmom}
\hat{G'}=\hbar(\Gamma_1\hat{X}_1+\Gamma_2\hat{X}_2+\kappa\hat{X}_3),
\end{equation}
of Hermitian generators
\begin{eqnarray}\label{gen123}
\hat{X}_1&=&\hat{A}_{s_1}\hat{A}_{i_1}+\hat{A}_{s_1}^{\dag}\hat{A}_
{i_1}^{\dag},\quad\hat{X}_2=\hat{A}_{s_2}\hat{A}_{i_2}+\hat{A}_{s_2}^{\dag}
\hat{A}_{i_2}^{\dag} \nonumber \\
&&\hat{X}_3=\hat{A}_{i_1}\hat{A}_{i_2}^
{\dag}+\hat{A}_{i_1}^{\dag}\hat{A}_{i_2}.
\end{eqnarray}
These, together with other three Hermitian generators,
\begin{eqnarray}\label{gen456}
\hat{X}_4&=&i(\hat{A}_{s_1}\hat{A}_{i_2}-\hat{A}_{s_1}^{\dag}\hat{A}_
{i_2}^{\dag}),\enspace\hat{X}_5=i(\hat{A}_{i_1}\hat{A}_{s_2}-\hat{A}_{i_1}
^{\dag}\hat{A}_{s_2}^{\dag})\nonumber \\
&&\hat{X}_6=\hat{A}_{s_1}\hat{A}_{s_2}^
{\dag}+\hat{A}_{s_1}^{\dag}\hat{A}_{s_2},
\end{eqnarray}
comprise a closed six dimensional algebra
\begin{equation}\label{alg}
[\hat{X}_i,\hat{X}_j]=iC_{ij}^k\hat{X}_k,\quad i,j,k=1,..,6.
\end{equation}
The real structure coefficients $C_{ij}^k$ need not be specified here;
note only that $[\hat{X}_1,\hat{X}_2]=[\hat{X}_3,\hat{X}_6]=[\hat{X}_4,
\hat{X}_5]=0$.
Closeness of the algebra (\ref{alg}) guarantees
the possibility to decompose the evolution operator of the system
as follows
\begin{eqnarray}\label{decomp}
e^{\frac{i}{\hbar}\hat{G}z}&=&e^{ig_5(z)\hat{X}_5}e^{ig_4(z)\hat{X}_4}
e^{ig_1(z)\hat{X}_1}e^{ig_2(z)\hat{X}_2}\nonumber \\
&& \times e^{ig_3(z)\hat{X}_3}
e^{ig_6(z)\hat{X}_6}.
\end{eqnarray}
Physically, this corresponds to replacing
the discussed experimental setup with a sequence of simpler
devices like downconvertors
($\hat{X}_1$, $\hat{X}_2$, $\hat{X}_4$,
$\hat{X}_5$), and beamsplitters ($\hat{X}_3$, $\hat{X}_6$)
generating the same output fields \cite{fiurasek}.
To a given interaction length $L$ of the original device
(\ref{strmom}) there corresponds a substituting scheme
(\ref{decomp}) with a particular choice of the coupling
parameters, $g_i=g_i(L)$.
The reasons for substituting the original device are twofold.
First, for the given values of the parameters $g_i$ the
input-output transformation provided by the substituting
scheme is relatively simple, thus suitable for physical discussion.
Moreover, well below threshold, where
all the output intensities are low, the parameters $g_1$,
$g_2$, $g_4$, and $g_5$ of the downconvertors of the corresponding
substituting scheme are expected to be small.
Then one can solve for the input-output
transformation to only a low order in those four
parameters resulting in further simplifications.
In this way the long-time dynamics of the original
device (Fig.~\ref{fig-coupler}) is mapped into
the short-time dynamics of the substituting scheme.

\begin{figure}
\centerline{\psfig{file=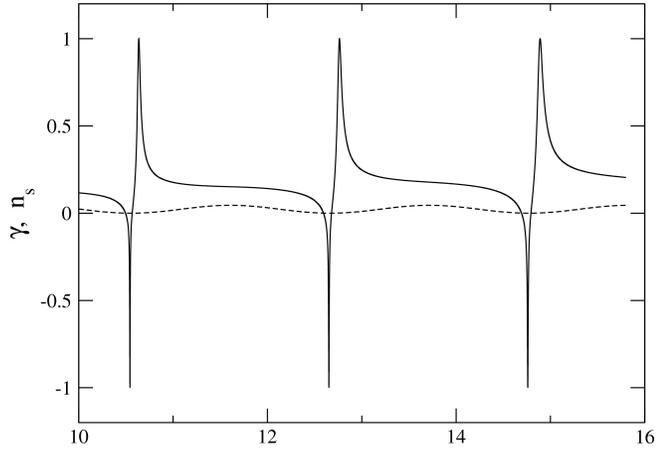,width=0.99\linewidth}}
\caption{Mutual coherence of signal beams (solid line);
mean total number of signal photons for the same interaction length
$L$ (dotted line); $\kappa=3$, $\Gamma_1=0.1$,  $\Gamma_2=0.3$.}
\label{fig-coherence}
\end{figure}

Since we are only interested in the spontaneous process
where all the modes of the original device start from vacuum
the presence of beamsplitters $\hat{X}_3$ and $\hat{X}_6$
in the substituting scheme is irrelevant [their exponential
operators act on vacuum].
This is why the particular ordering in (\ref{decomp})
has been chosen. The remaining four downconvertors,
see Fig.~\ref{fig-substit}, constitute the sought
substituting scheme which, for input vacuum states, is equivalent
to the original setup (Fig.~\ref{fig-coupler}).
Here the links between the experiment on the induced emission
\cite{mandel} and our continuously coupled device
start to appear. Two simultaneous experimental
arrangements of Zou {\em et al.}\/ can be recognized
in Fig.~\ref{fig-substit}; one being formed by downconvertors
$1$ and $5$, the other being formed by the remaining two.
This similarity will help us to explain the induced
coherence observed in Fig.~\ref{fig-coherence}.

In order to keep algebra simple we will assume that
at most one downconverted pair of photons is present
in the substituting scheme.
This is usually a good approximation under standard
experimental conditions. This also corresponds to the situation when
the original device operates well below threshold, see
Fig.~\ref{fig-coherence}. Low mean number of output signal photons
implies small magnitudes of nonlinearities $|g_i|$,
as can be seen from the following inequality,
\begin{eqnarray} \label{signal-tot}
\langle\hat{n}_{s_1}\rangle+ \langle\hat{n}_{s_2}\rangle&
\ge&\sinh^2g_1+\sinh^2g_2 +
\sinh^2g_4\nonumber\\
&&+\sinh^2g_5,
\end{eqnarray}
which immediately yields an upper bound on $|g_i|$,
\begin{equation} \label{small}
|g_i|\le\sinh^{-1}(\sqrt{\langle\hat{n}_{s_1}\rangle+
\langle\hat{n}_{s_2}\rangle}).
\end{equation}

The evolution operator of the first pair
of downconvertors in Fig.~\ref{fig-substit}
to the first order in the nonlinear coupling parameters reads
\begin{equation} \label{prvni-dvojice}
U_1=1+ig_1\hat{A}^{\dag}_{s_1}\hat{A}^{\dag}_{i_1}+
ig_2\hat{A}^{\dag}_{s_2}\hat{A}^{\dag}_{i_2}.
\end{equation}
Similarly, the evolution of the second pair of downconvertors
is governed by the operator
\begin{equation} \label{druha-dvojice}
U_2=1+g_4\hat{A}^{\dag}_{s_1}\hat{A}^{\dag}_{i_2}+
g_5\hat{A}^{\dag}_{s_2}\hat{A}^{\dag}_{i_1}.
\end{equation}
Note that annihilation operators do not contribute
to Eqs.~(\ref{prvni-dvojice}) and (\ref{druha-dvojice})
because we assume that at most one pair
of photons is created in the substituting scheme.

The input vacuum state develops into the output state
\begin{eqnarray}\label{stav}
|\Psi\rangle&\approx& (1-|{\rm vac}\rangle\langle {\rm vac}|)
U_2U_1|{\rm vac}\rangle\approx
g_4|1001\rangle+g_5|0110\rangle\nonumber \\
&&+ig_1|1100\rangle+ig_2|0011\rangle,
\end{eqnarray}
where the redundant vacuum component has been projected out and
only terms to the first order in the nonlinear coupling parameters
have been kept. We have introduced a short-hand notation
$|n_1n_2n_3n_4\rangle\equiv |n_1\rangle_{s_1}|n_2\rangle_{i_1}|n_3
\rangle_{s_2}|n_4\rangle_{i_2}$.
Here and in the following we omit unimportant normalization
factors.
The mutual coherence of the
signal beams is, according to Eq.~(\ref{coherence}), given by the
transition probability amplitude between the states
\begin{eqnarray} \label{annih}
\hat{A}_{s_2}|\Psi\rangle&=&
g_5|0100\rangle+ig_2|0001\rangle,
\nonumber\\
\hat{A}_{s_1}|\Psi\rangle&=&
ig_1|0100\rangle+g_4|0001\rangle.
\end{eqnarray}
It is convenient to define two vectors whose components
are proportional to the nonlinear coupling parameters of the four
downconvertors,
\begin{equation} \label{vektory}
{\bf u}=\left(
\begin{array}{c}
g_1\\ g_4 \\0
\end{array}\right),
\qquad
{\bf v}=\left(
\begin{array}{c}
g_5\\ -g_2 \\0
\end{array}\right).
\end{equation}
In terms of vectors ${\bf u}$ and ${\bf v}$, the normalized mutual
coherence of signal beams reads
\begin{equation} \label{gamma}
|\gamma|^2=\frac{|{\bf u}\cdot{\bf v}|^2}{u^2 v^2}=
\frac{(g_1g_5-g_2g_4)^2}{(g_1^2+g_4^2)(g_2^2+g_5^2)}.
\end{equation}
The calculation of the mutual coherence of signal beams
thus has been reduced to considering the mutual geometry of the
vectors ${\bf u}$ and
${\bf v}$.
If the parameters of the substituting scheme are such
that the vectors are orthogonal, the mutual coherence
$\gamma$ vanishes and the signal beams do not interfere.
In the opposite extreme case when the vectors ${\bf u}$ and
${\bf v}$
are almost collinear, the mutual coherence attains its
maximum magnitude and the signal beams become first-order
coherent. An example of the latter situation is the
experiment of Zou {\em et al.}\/ \cite{mandel}, which is
a special case of the setup in Fig.~\ref{fig-substit}
with $g_1=g_5=0$ or $g_2=g_4=0$.

In the following we will show that the mutual coherence of signal
beams strongly depends on the amount of the which-way information
about them that is carried by the idler beams.
Before any measurement on the system is attempted,
the only which-way information about the signal photon
that is available arises from different
intensities of signal beams. This information will be called
prior which-way information. Provided the idler beams
carry information about the signal photon, the prior
which-way information can be updated by performing
a suitable measurement on them.

First let us discuss the incoherent case.
When the vectors ${\bf u}$ and ${\bf v}$ are orthogonal,
the output state $|\Psi\rangle$ can be rewritten as follows,
\begin{equation} \label{tensor}
|\Psi\rangle=
\frac{g_2}{g_1}|\varphi_1\rangle_i
|01\rangle_s+|\varphi_2\rangle_i
|10\rangle_s,\quad ({\bf u}\perp{\bf v})
\end{equation}
where the states $|\varphi_1\rangle$ and $|\varphi_2\rangle$ of
the idler beams,
\begin{eqnarray} \label{meas}
|\varphi_1\rangle&=&
g_4|10\rangle_i+ig_1|01\rangle_i,\nonumber \\
|\varphi_2\rangle&=& ig_1|10\rangle_i+g_4|01\rangle_i,
\end{eqnarray}
are orthogonal. This means that there is a measurement on the
idler beams yielding a perfect which-way information about the
signal photon, for instance, the measurement having the spectral
decomposition
\begin{equation}\label{hermit}
\hat{S}=|\varphi_1\rangle\langle\varphi_1|-
|\varphi_2\rangle\langle\varphi_2|.
\end{equation}
If the outcome corresponding to $|\varphi_1\rangle$ is detected
the signal photon is projected to mode $s_2$. If, on the other hand,
$|\varphi_2\rangle$ is detected, the signal
photon is projected to mode $s_1$.
Perfect knowledge of the signal photon's
path precludes the interference.
Provided the parameters of the substituting scheme are such that
the vectors ${\bf u}$ and ${\bf v}$ are not completely orthogonal,
the decomposition (\ref{tensor}) with orthogonal states of the
idler beams is not possible; the knowledge of the signal photon's
path is then only partial.
Nevertheless, one can still
think of the measuring apparatus (\ref{hermit}) as
of the ideal which-way apparatus -- ideal in the sense that
it gives perfect which-way knowledge in the limit
${\bf u}\cdot {\bf v}\rightarrow 0$.
Eigenvectors of $\hat{S}$ can be used to decompose the output state
$|\Psi\rangle$ in the spirit of (\ref{tensor}),
\begin{eqnarray}\label{tensor_gen}
|\Psi\rangle&=&|{\bf u}\times{\bf v}|\,|\varphi_1\rangle_i
|01\rangle_s\nonumber\\
&+&|\varphi_2\rangle_i(-i{\bf u}\cdot{\bf v}|01\rangle_s+
{\bf u}^2|10\rangle_s).
\end{eqnarray}
Now, the gained amount of information about the signal photon
will depend on the outcome of the measurement of $\hat{S}$.
If the result $|\varphi_1\rangle$ is detected,
the signal photon will be localized in the mode $s_2$ with
certainty, and therefore it will not contribute to the interference pattern.
If, however, we end up with the result $|\varphi_2\rangle$,
the state of the signal photon will become a coherent superposition
of states ``photon being in mode $s_1$'' and ``photon being in
mode $s_2$'', and $|\gamma|$ attains the maximum value of
one. Since both the outcomes occur in random, the mutual
coherence $|\gamma|$ becomes a weighted sum of zero and one.
As the angle between the vectors ${\bf u}$ and ${\bf v}$ gets smaller,
the probability of the former unambiguous case to happen
gets smaller, too. When the vectors become collinear,
the output state (\ref{stav}) becomes a factorized state.
In this case, the state of the signal field after
the measurement will be the pure superposition
\begin{equation} \label{superp}
|\Psi\rangle_s= \pm i v|01\rangle_s+
u|10\rangle_s,\qquad ({\bf u}\,\|\,{\bf v}),
\end{equation}
and the mutual coherence will be maximum, in agreement with
Eq.~(\ref{gamma}). Because the diagonal elements of
the state (\ref{superp}) in $\{|10\rangle_s,|01\rangle_s\}$
basis are the same as those of the pre-measurement
state (\ref{stav}), the posterior which-way information
equals the prior which-way information and no additional
which-way information is gained by the measurement.

It is interesting to look closely at the measurement $\hat{S}$ and
its relation to the optimum which-way measurement in the
experiment of Zou {\em et al.}\/ which is nothing else than the
counting of idler photons in modes $i_1$ and $i_2$. Eigenvectors of
$\hat{S}$ (\ref{meas}) can be parameterized by an angle $\phi$,
\begin{eqnarray}\label{spin}
|\varphi_1\rangle=\sin\phi|10\rangle_i
+i\cos\phi|01\rangle_i,\nonumber\\
|\varphi_2\rangle=i\cos\phi|10\rangle_i
+\sin\phi|01\rangle_i,
\end{eqnarray}
\begin{equation} \label{angle}
\sin\phi=\frac{g_4}{\sqrt{g_1^2+g_4^2}}.
\end{equation}
Hence the measurement $\hat{S}$
can be rewritten in terms of the $z$-component of the
Stokes operator acting on the two-mode idler field,
$\hat{S}_z=|01\rangle\langle01|-|10\rangle\langle10|$,
\begin{equation} \label{stokes}
\hat{S}=e^{-i\hat{\sigma}_y\phi}\hat{S}_z
e^{i\hat{\sigma}_y\phi},
\end{equation}
where $\hat{\sigma}_y$ is Pauli matrix.
Notice that on the subspace of the Hilbert space of the idler modes
with the total number of idler photons  $n_{i_1}+n_{i_2}=1$ the
operator $\hat{S}_z$ is just the
measurement of the difference of the numbers of photons in modes
$i_2$ and $i_1$, $\hat{S}_z=\hat{n}_{i_2}-\hat{n}_{i_1}$.
This is the ideal which-way measurement in the experiments of
Zou {\em et al.}\/ Here the situation is similar. The
ideal which-way detection differs from it only by a rotation given
by the coupling parameters of the substituting scheme.

The parameters of the substituting scheme can be found
by differentiating both sides of Eq.~(\ref{decomp})
and rearranging the right-hand side
\cite{algebra}.
In this way one obtains a system of nonlinear
differential equations for the sought parameters,
which can be solved by numerical integration.
There is also another approach \cite{fiurasek} which makes use
of correlation functions and the fact that all the input modes
are in the vacuum state. This procedure yields explicit
expressions for the unknown parameters $g_i$, see Appendix.


The oriented angle between the vectors ${\bf u}$ and ${\bf v}$
characterizing the substituting  scheme is displayed in
Fig.~\ref{fig-param} for the same coupling strengths
of the original device as that in Fig.~\ref{fig-coherence}.
Notice that zeros/maxima of the
signal mutual coherence in Fig.~\ref{fig-coherence} indeed coincide
with the interaction lengths for which the vectors ${\bf u}$ and
${\bf v}$ become orthogonal/parallel, in agreement with Eq.~(\ref{gamma}).
One can also check that all the coupling parameters of the active elements
are small; in this case it holds that $|g_{1,2,4,5}|\le 0.2$.
Since the probability of having more than one photon pair in the
substituting scheme scales as $\sum_i g_i^4$, the approximations
(\ref{prvni-dvojice}) and (\ref{druha-dvojice}) are fully justified
well below threshold.
This proves our claim that also in the case of continuously
coupled downconvertors the coherence of signal beams is
governed by the principal indistinguishability
of signal photons. Our device can thus be regarded as an
interesting generalization of the experimental setup of
Zou {\em et al.}\/ where the
distinguishability of signal photons can be controlled by the
length of the device itself rather than by using auxiliary
optical elements.

\subsection{Setup of Ou {\em et al.}\/}

An alternative representation of two continuously
coupled downconvertors is possible.
It has been shown in \cite{braunstein}
that any four-mode unitary transformation
leaving the set of Gaussian states invariant can be realized
by using only four single-mode squeezers placed in between two
four-port linear interferometers. In our case we can further simplify
such scheme if we use  two-mode squeezers (nondegenerate downconvertors)
instead of single-mode squeezers.
The continuously coupled downconvertors can be replaced with a
sequence of two beamsplitters followed by two downconvertors and another
two beamsplitters. Since we consider spontaneous process,
the first pair of beamsplitters can be omitted, see
Fig.~\ref{fig-substit-mirek}.
Notice that actually this is the same arrangement as the well-known
two-photon interferometer of Ou {\em et al.}\/ \cite{ou}.
Its four relevant parameters, the nonlinear coupling
parameters $g_1$ and $g_2$ of the two downconvertors, and
the mixing angles $\phi_s$, $\phi_i$ of the two
beamsplitters are determined by the coupling strengths
$\Gamma_1$, $\Gamma_2$, $\kappa$, and the interaction length
$L$ of the original
continuously coupled device.

The mutual coherence of signal beams can now be expressed  as
\begin{equation}
\gamma=\frac{\langle \hat{n}_{s_1}\rangle
-\langle \hat{n}_{s_2}\rangle }
{\sqrt{4\langle \hat{N}_{s_1}\rangle \langle \hat{N}_{s_2}\rangle}}
 \sin(2\phi_s),
 \label{gamma-mirek}
\end{equation}
where
\begin{eqnarray}
\langle \hat{N}_{s_1}\rangle &=&
\langle \hat{n}_{s_1}\rangle\cos^2\phi_s+
\langle \hat{n}_{s_2}\rangle\sin^2\phi_s,
\nonumber \\
\langle \hat{N}_{s_2}\rangle &=&
\langle \hat{n}_{s_1}\rangle\sin^2\phi_s+
\langle \hat{n}_{s_2}\rangle\cos^2\phi_s,
\end{eqnarray}
and
\begin{equation}
\langle \hat{n}_{s_j}\rangle=\sinh^2 g_j
\label{ntilde}
\end{equation}
are mean numbers of  photons in signal modes
at the output of the two downconvertors.
The degree of coherence $\gamma$ depends on the mixing angle $\phi_{s}$
of the signal beamsplitter. However, more important is the dependence
on the {\em difference} of the mean numbers of  photons
$\langle \hat{n}_{s_j}\rangle$. If $g_1=g_2$, then
$\langle \hat{n}_{s_1}\rangle=\langle \hat{n}_{s_2}\rangle$
and $\gamma=0$. On the other hand, if $g_1=0$ or $g_2=0$
then $|\gamma|=1$ irrespective of $\phi_{s}$ and the two output
signal modes are fully coherent.
If the two continuously coupled downconvertors effectively behave like
a single downconvertor followed by beamsplitters,
then the two output signal modes stem from a single mode in chaotic state
mixed with vacuum at a beamsplitter. This leads to the full
coherence of the two output signal modes. On the other hand,
if the device effectively behaves like two identical
downconvertors then both signal modes mixed at BS$_s$
are in chaotic state with the same mean numbers of photons
$\langle \hat{n}_{s_1}\rangle=\langle \hat{n}_{s_2}\rangle$
and no coherence can be observed at the output.
When the interaction length $L$ is varied one can continuously
move from the regime where one downconvertor is dominant to
the regime where both downconvertors have comparable output.
These transitions result in sharp peaks observed in
Fig.~\ref{fig-coherence}.
A~typical behavior of parameters $g_1$ and $g_2$ is
shown in Fig.~\ref{fig-g1g2}. Notice that interaction lengths
for which one of the parameters is zero or $|g_1|=|g_2|$
correspond to full or no coherence in Fig.~\ref{fig-coherence}.

The analysis of the coherence properties of continuously
coupled downconvertors in terms of which-way information presented
in the previous subsection can equivalently be done using the
present scheme. Since the analysis would lead to the same
interpretation as in previous subsection we do not repeat it here.
We just note that it follows from Eq. (\ref{ntilde}) that
the parameters $g_1$ and $g_2$ of the present
substituting scheme are small well below the threshold
where the output intensity is low, see also Fig.~\ref{fig-g1g2}.

Having seen that one device
-- two continuously coupled downconvertors --
can be replaced with two different schemes
resembling two famous experiments  one may ask
whether the experimental arrangements themselves are in
some sense equivalent.
The affirmative answer of course stems from the fact that
both the experiments \cite{mandel} and \cite{ou}
utilize the same optical elements whose action falls
to the same group of transformations.
Just for completeness let us show how the two-photon
interferometer (Fig.~\ref{fig-substit-mirek})
can be used to analyze
the experiment on the induced coherence without induced emission
\cite{mandel}. The setup
consists of two  downconvertors with parameters $r_1$ and $r_2$.
The idler mode coming from the first crystal is partially injected in the
second one. This is realized, e.g. by placing a beamsplitter with mixing
angle $\psi$ in between them. The input-output
transformation of this device reads \cite{rehacek96}
\begin{eqnarray}
\hat{A}_{s_1,\rm out}&=&\hat{A}_{s_1}\cosh r_1 + i \hat{A}_{i_1}^\dagger
 \sinh r_1,
\nonumber \\
\hat{A}_{s_2,\rm out}&=&-i \hat{A}_{s_1}\sinh r_1 \sin \psi \sinh r_2
+\hat{A}_{s_2}\cosh r_2
\nonumber \\ & &
+\hat{A}_{i_1}^\dagger \cosh r_1 \sin \psi \sinh r_2
+i \hat{A}_{i_2}^\dagger \cos \psi \sinh r_2,
\nonumber \\
\hat{A}_{i_1,\rm out}&=& i \hat{A}_{s_1}^\dagger \sinh r_1 \cos \psi
\nonumber \\ & &
+\hat{A}_{i_1}\cosh r_1 \cos \psi + i \hat{A}_{i_2}\sin\psi,
\nonumber \\
\hat{A}_{i_2,\rm out} &=&
 -\hat{A}_{s_1}^\dagger  \sinh r_1 \sin\psi \cosh r_2
 +i \hat{A}_{s_2}^\dagger \sinh r_2
 \nonumber \\ & &
 +i \hat{A}_{i_1}\cosh r_1 \sin \psi \cosh r_2
 +\hat{A}_{i_2}\cos \psi \cosh r_2.
\nonumber \\
\label{mandelio}
\end{eqnarray}
In the original setup of Zou {\em et al.}\/ the mutual coherence of the
signal beams was  controlled by the
value of mixing angle $\psi$. When the idler beams were perfectly
superposed, $\psi=\pi/2$, the signal beams were fully coherent, $|\gamma|=1$,
because the idler photons did not contain any which-way information.
In the opposite extreme $\psi=0$, the signal modes were
completely incoherent, $\gamma=0$, because the idler photons
in principle could allow one to determine whether the signal photon
had come from the first or the second downconvertor.

Let us now show how this dependence of $\gamma$ on $\psi$
is reflected in the substituting scheme shown in Fig.~\ref{fig-substit-mirek}.
In Fig.~\ref{fig-equiv} we plot $\gamma$ and $g_1$, $g_2$ as
functions of the mixing
angle $\psi$. As could be expected,
$\gamma$ increases with $\psi$
from $0$ at $\psi=0$ to $1$ at $\psi=\pi/2$.
Also the squeezing constant $g_1$ increases with $\psi$,
while $g_2$ decreases and reaches $0$ at $\psi=\pi/2$.
The coherence $\gamma$ increases with growing ratio
$g_1/g_2$ in agreement with Eq. (\ref{gamma-mirek}).
If the idler modes of the two downconvertors are perfectly aligned,
then the whole setup is equivalent to a
single downconvertor followed by a beamsplitter.
Since there is only one source of signal photons in this case,
the output signal beams are fully coherent.

\section{Discussion and conclusions} \label{zeno}

We have found that idler beams of two continuously coupled
downconvertors become periodically strongly correlated
(uncorrelated) with signals in the course of evolution.
This is reflected in the entangled (disentangled) nature of
the signal-idler field (\ref{tensor}) and resembles
the interaction of a quantum system (signal modes) with a
quantum meter (idler modes). One may loosly say that the signal
modes are periodically ``observed'' by the idler modes.
Yet one has to realize
that no {\em bona fide} measurement is being performed in this way.
An essential part of any such measurement -- the reading of the
quantum meter by a
classical apparatus -- is missing here.
The signal field is not projected
but continues the unitary evolution.
Nevertheless,  the entanglement of signal and idler fields,
which is responsible for destroying the signal coherence
is enough to disturb the phase relations between
interacting modes and slow down the downconversion process
\cite{luis}.
As the coupling of idler modes becomes stronger the
frequency of ``observations'' of the signal field by the idler
one increases. In the limit of very large $\kappa$
the two downconversion processes become completely frozen
and no photon pairs are being created anymore.
This effect, usually called the quantum Zeno effect \cite{others},
provides an alternative explanation of why two strongly
coupled downcorvertors operate below threshold.
Similar situation in another experimental setup
has been analyzed in \cite{reh-zeno}. This effect has been
interpreted as the quantum Zeno effect caused by a continuous
observation of one mode of light by another one.
It is well known that frequently repeated measurements
and continuous observation (or interactions)
can hinder evolution of a quantum system in a similar way
\cite{zeno-cont}.
Some quantitative statements relating
the frequency of discrete measurements to
the strength of continuous interactions having the same
effect can even be found in the literature \cite{zeno-relat}.
The experimental arrangement discussed in this paper
is an interesting example of a system
where such discrete disturbances naturally arise from
a continuous interaction between its constituent
parts.

\section*{Acknowledgement}
We would like to thank A. Luis and R. Filip for helpful comments.
This work was supported by the project
LN00A015 of the Czech Ministry of Education.

\appendix
\section*{Parameters of the  substituting schemes}
This derivation closely follows the general
procedure of Ref. \cite{fiurasek}, Sec.~IIC.
In the Heisenberg picture, the vector of operators
\begin{equation}
{\bf \hat{A}}=
\left(
\begin{array}{c}
\hat{A}_{s_1} \\
\hat{A}_{s_2}  \\
\hat{A}_{i_1}^\dagger \\
\hat{A}_{i_2}^\dagger
\end{array}
\right)
\end{equation}
transforms according to
\begin{equation}
{\bf \hat{A}}_{\rm out}= {\bf M} {\bf \hat{A}}_{\rm in},
\label{Ainout}
\end{equation}
where ${\bf M}=\exp(i{\bf H} L)$ and
\begin{equation}
{\bf H}=
\left(
\begin{array}{cccc}
0 & 0 & \Gamma_1 & 0 \\
0 & 0 & 0 & \Gamma_2 \\
-\Gamma_1 & 0 & 0 & -\kappa \\
0 & -\Gamma_2 & -\kappa & 0
\end{array}
\right).
\end{equation}

\begin{figure}
\centerline{\psfig{file=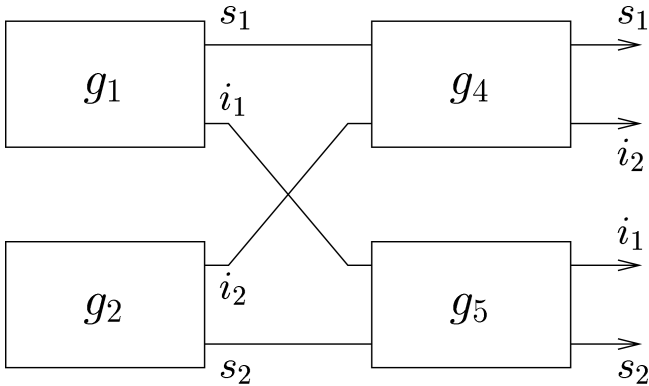,width=0.99\linewidth}}
\vspace{0.5cm}
\caption{Substituting scheme for the two continuously
coupled downconvertors shown in Fig. \protect{\ref{fig-coupler}}. Notice
that the continuous interaction of three modes is replaced by a
sequence of simpler devices at the expense of doubling the number
of active elements.
}\label{fig-substit}
\end{figure}

\begin{figure}
\centerline{\psfig{file=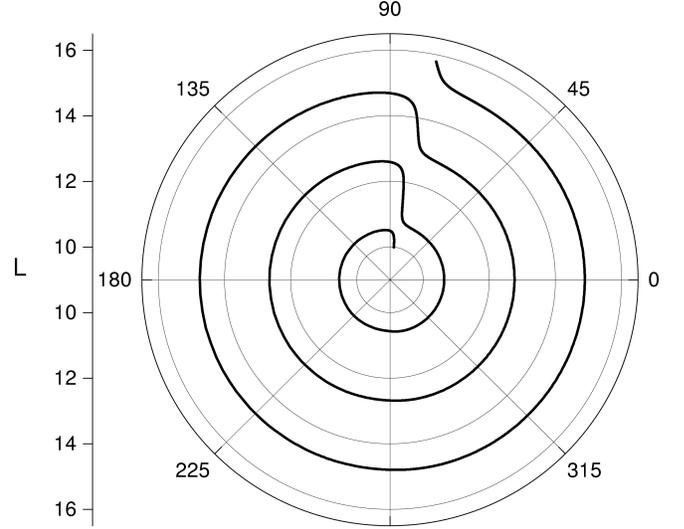,width=0.99\linewidth}}
\caption{Mutual orientation of vectors ${\bf u}$ and ${\bf v}$
characterizing the substituting scheme of
Fig.~\protect{\ref{fig-substit}}
which can substitute the continuously coupled device
with $\kappa=3$, $\Gamma_1=0.1$, and $\Gamma_2=0.3$ (compare to
Fig.~\protect{\ref{fig-coherence}}).}
\label{fig-param}
\end{figure}

\begin{figure}
\centerline{\psfig{file=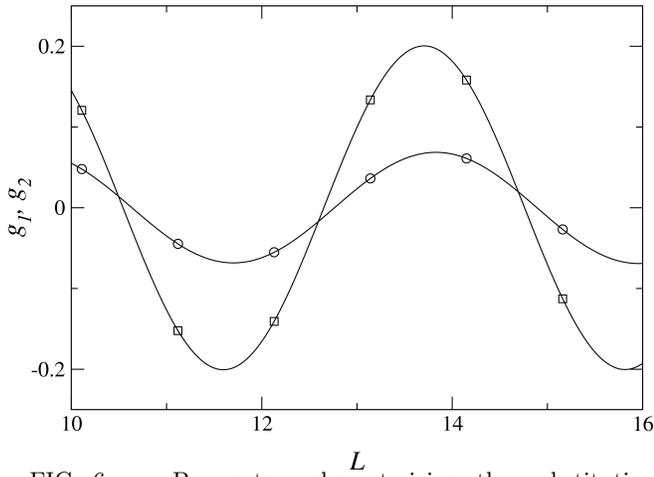,width=0.99\linewidth}}
\caption{
Parameters characterizing the substituting scheme
(two-photon interferometer of Ou {\em et al.}\/) for
two continuously coupled downconvertors.
The coupling strengths of the
original device are the same as in Figs.~\protect{\ref{fig-coherence}}
and \protect{\ref{fig-param}}; $g_1$ (circles), $g_2$ (squares).}
\label{fig-g1g2}
\end{figure}

In order to find analytical expressions for the parameters of
the substituting schemes, we consider output state generated from the
vacuum and propagate this state backward through the schemes. At each step,
some correlations between the four modes vanish, which provides us with
equations for the unknown parameters. Let us begin with the scheme shown
in Fig. \ref{fig-substit}.  When we propagate the output modes
in front of the second pair of downconvertors, we get
 \[
 {\bf \hat{A}}_{\rm c}= {\bf M}_{45} {\bf \hat{A}}_{\rm out},
 \]
  where
\begin{equation}
{\bf M}_{45}=
\left(
\begin{array}{cccc}
\cosh g_4 & 0 & 0 & -\sinh g_4 \\
0 & \cosh g_5 & -\sinh g_5 & 0 \\
0 & -\sinh g_5 & \cosh g_5 & 0 \\
-\sinh g_4 & 0 & 0 & \cosh g_4
\end{array}
\right).
\end{equation}
The modes $s_{1,c}$ and $i_{1,c}$ are not correlated
with $s_{2,c}$ and $i_{2,c}$. In particular,
\begin{eqnarray}
\langle \Delta\hat{A}_{s_1,\rm c} \Delta\hat{A}_{i_2, \rm c}\rangle =0,
\qquad
\langle \Delta\hat{A}_{s_2,\rm c} \Delta\hat{A}_{i_1,\rm c}\rangle =0.
\label{SIC}
\end{eqnarray}
We can solve these two  equations for the parameters $g_4$ and $g_5$,
\begin{eqnarray}
\tanh (2 g_4)=\frac{2D_{s_1 i_2, \rm out}}{B_{s_1,\rm out}
+ B_{i_2, \rm out} + 1},
\nonumber  \\
\tanh (2 g_5)=\frac{2D_{s_2 i_1, \rm out}}{B_{s_2,\rm out}
+ B_{i_1, \rm out} + 1}.
\label{g45exp}
\end{eqnarray}

where
\begin{eqnarray}
D_{jk,\rm out}&=&\langle \Delta \hat{A}_{j,\rm out} \Delta
\hat{A}_{k,\rm out}\rangle,
\nonumber \\
B_{j,\rm out}&=&\langle \Delta \hat{A}_{j,\rm out}^\dagger \Delta
\hat{A}_{j,\rm out}\rangle.
\label{BDout}
\end{eqnarray}
Since all the input modes are in vacuum state it is easy to express the
correlations (\ref{BDout}) in terms of the elements of matrix
$\bf M$ \cite{fiurasek}. We find,
\begin{eqnarray}
D_{s_1 i_1, \rm out}&=& M_{11} M_{31}^\ast+M_{12} M_{32}^\ast,
\nonumber \\
D_{s_1 i_2, \rm out}&=& M_{11} M_{41}^\ast+M_{12} M_{42}^\ast,
\nonumber \\
D_{s_2 i_1, \rm out}&=& M_{21} M_{31}^\ast+M_{22} M_{32}^\ast,
\nonumber \\
D_{s_2 i_2, \rm out}&=& M_{21} M_{41}^\ast+M_{22} M_{42}^\ast,
\nonumber \\
B_{s_1,\rm out}&=&|M_{13}|^2+|M_{14}|^2,
\nonumber \\
B_{s_2,\rm out}&=&|M_{23}|^2+|M_{24}|^2,
\nonumber \\
B_{i_1,\rm out}&=&|M_{31}|^2+|M_{32}|^2,
\nonumber \\
B_{i_2,\rm out}&=&|M_{41}|^2+|M_{42}|^2.
\label{DBM}
\end{eqnarray}

In the next step we propagate ${\bf \hat{A}}_{\rm c}$ back in front of the
first pair of downconvertors,
${\bf \hat{A}}_{\rm in}={\bf M}_{12} {\bf \hat{A}}_{\rm c}$,
\begin{equation}
{\bf M}_{12}=
\left(
\begin{array}{cccc}
\cosh g_1 & 0 & -i\sinh g_1 & 0 \\
0 & \cosh g_2 & 0 & -i\sinh g_2 \\
i \sinh g_1 & 0 & \cosh g_1 & 0 \\
0 & i\sinh g_2 & 0 & \cosh g_2
\end{array}
\right).
\end{equation}
Since the input modes are in vacuum states, all correlations should
vanish. In particular, we have
\begin{eqnarray}
\langle \Delta\hat{A}_{s_1,\rm in} \Delta\hat{A}_{i_1, \rm in}\rangle =0,
\qquad
\langle \Delta\hat{A}_{s_2,\rm in} \Delta\hat{A}_{i_2,\rm in}\rangle =0.
\label{corr12}
\end{eqnarray}
Upon solving the Eqs. (\ref{corr12}) for the parameters $g_1$ and $g_2$,
we get
\begin{eqnarray}
\tanh (2 g_1)=\frac{-2iD_{s_1 i_1, \rm c}}{B_{s_1,\rm c}+ B_{i_1, \rm c} + 1},
\nonumber  \\
\tanh (2 g_2)=\frac{-2iD_{s_2 i_2, \rm c}}{B_{s_2,\rm c}+ B_{i_2, \rm c}
+ 1}.
\label{g12exp}
\end{eqnarray}
The correlations $D_{jk, \rm c}$ and $B_{j, \rm c}$  of operators
$\hat{A}_{j,\rm c}$ can be expressed in terms of the elements of the matrix
${\bf M}_{\rm c}= {\bf M}_{45} {\bf M}$. One can directly use Eq. (\ref{DBM})
where $\bf M$ is replaced with ${\bf M}_{\rm c}$. We note that
$D_{s_1 i_1,\rm c}$ and $D_{s_2 i_2, \rm c}$ are purely imaginary
hence the right-hand sides of Eq. (\ref{g12exp}) are real.

\begin{figure}
\centerline{\psfig{file=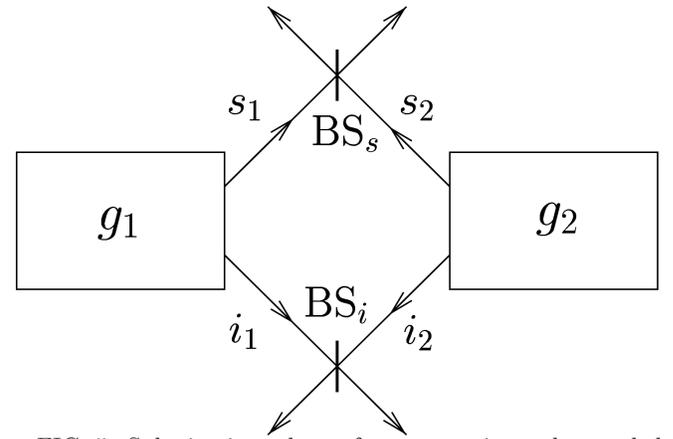,width=0.99\linewidth}}
\caption{Substituting scheme for two continuously coupled
downconvertors which is equivalent to the interferometer of
Ou {\em et al.}\/ Provided that all the input modes are in
vacuum state, two downconvertors and two beamsplitters
are enough to replace the original continuously coupled
device.}
\label{fig-substit-mirek}
\end{figure}

\begin{figure}
\centerline{\psfig{file=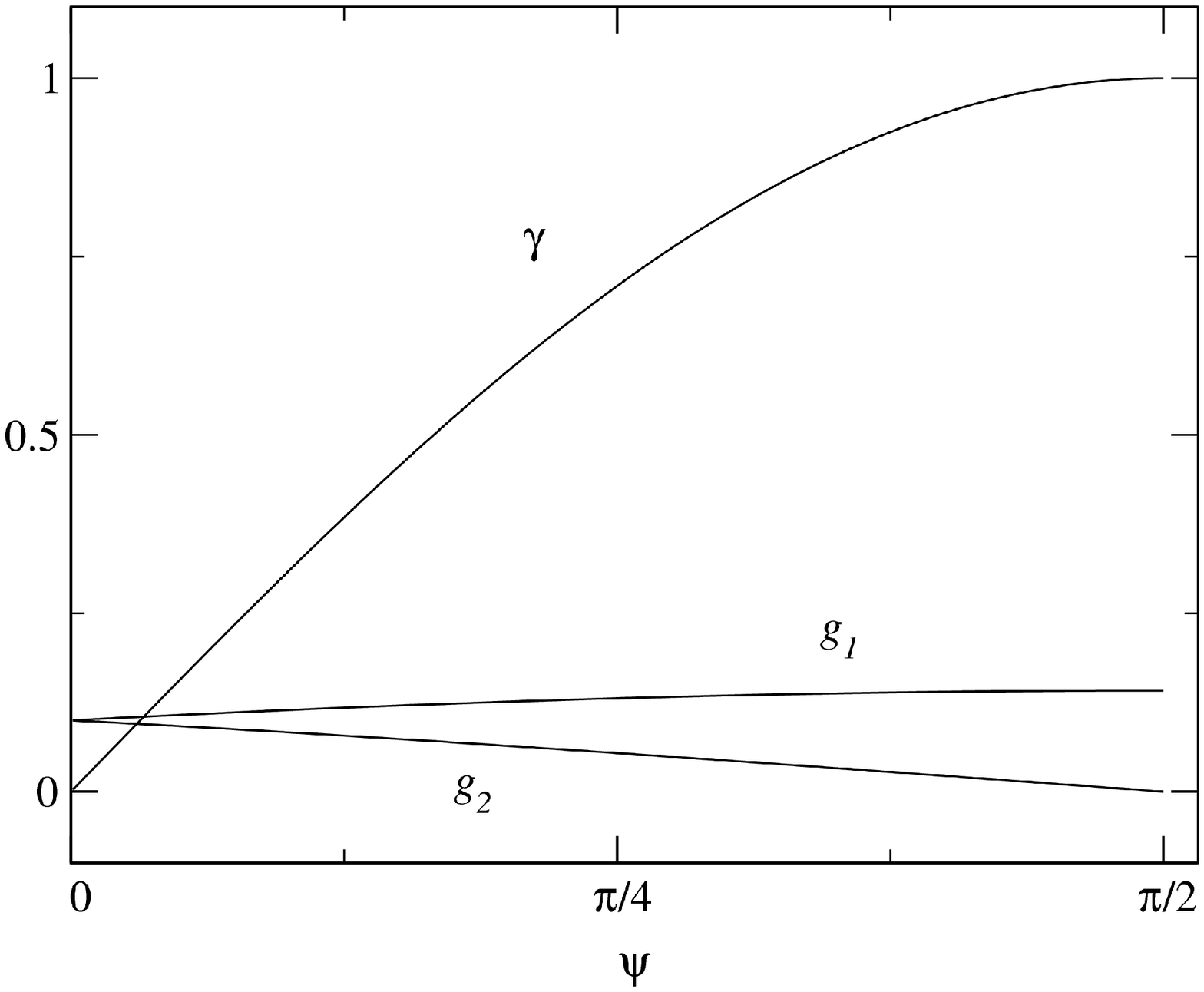,width=0.99\linewidth}}
\caption{
The two-photon interferometer of Ou {\em et al.}\/
(Fig.~\protect{\ref{fig-substit-mirek}}) can simulate the experimental
setup of Zou {\em et al.}\/ (two downconvertors with aligned idler beams).
The parameters characterizing the substituting scheme
are shown as functions of the mixing angle
of the beamsplitter controlling the alignment of idler beams
in the experiment of Zou {\em et al.}\/ ($\psi=\pi/2$ means perfect
alignment); $r_1=r_2=0.1$.}
\label{fig-equiv}
\end{figure}

The calculation of the parameters of the substituting scheme shown in Fig.
\ref{fig-substit-mirek} proceeds along similar lines.
We propagate the output modes in front of the beamsplitters,
${\bf \hat{A}}_{\rm c}= {\bf M}_{\phi} {\bf M} {\bf \hat{A}}_{\rm in}$,
where
\begin{equation}
{\bf M}_{\phi}=\left(
\begin{array}{cccc}
\cos\phi_s & -i\sin\phi_s & 0 & 0 \\
-i\sin\phi_s & \cos\phi_s & 0 & 0 \\
0 & 0 & \cos\phi_i & i\sin \phi_i  \\
0 & 0 & i\sin\phi_i & \cos \phi_i
\end{array}
\right).
\end{equation}
The conditions (\ref{SIC}) provide system of two coupled equations
for $\tan \phi_{s}$ and $\tan \phi_i$ whose solution reads
\begin{eqnarray}
\tan \phi_{s}&=&\frac{D_{s_1 i_2,\rm out}
    -iD_{s_1 i_1,\rm out}\tan\phi_i}
{D_{s_2 i_1,\rm out}\tan\phi_i+iD_{s_2 i_2,\rm out}},
\nonumber \\
\tan \phi_{i}&=&\frac{1}{2}(-p\pm \sqrt{p^2+4}),
\label{fiSI}
\end{eqnarray}
where
\begin{equation}
p=i\frac{D_{s_1 i_1,\rm out}^2+D_{s_1 i_2,\rm out}^2
  -D_{s_2 i_1,\rm out}^2-D_{s_2 i_2,\rm out}^2}
  {D_{s_1 i_1,\rm out}D_{s_1 i_2,\rm out}
  -D_{s_2 i_1,\rm out}D_{s_2 i_2,\rm out}}.
\end{equation}
When we know $\phi_{s,i}$ we calculate the matrix
${\bf M}_{\rm c}={\bf M}_{\phi}{\bf M}$ and determine
the parameters $g_1$ and $g_2$ from Eq. (\ref{g12exp}).

The above formulas  also enable us to find the
parameters of the interferometer of Ou {\em et al.}\/ \cite{ou}
which is equivalent to the  particular experimental
setup of Zou {\em et al.}\/ \cite{mandel}. The elements of
the appropriate matrix  $\bf M$ can be read off
the formulas (\ref{mandelio}) and then we can directly
use the Eqs. (\ref{g12exp}) and (\ref{fiSI}).

\newpage

\begin{thebibliography}{99}
\bibitem{mandel} X.\  Y.\  Zou, L.\  J.\  Wang and L.\  Mandel,
Phys.\   Rev.\  Lett.\  {\bf 67}, 318 (1991); L.\  J.\  Wang,
X.\  Y.\  Zou and L.\  Mandel,
Phys.\  Rev.\ A~{\bf 44}, 4614 (1991).
\bibitem{couplers}  G.\  Assanto ,
A.\  Laureti-Palma, C.\  Sibilia and
M.\  Bertolotti, Opt.\  Commun.\  {\bf 110}, 599 (1994);
J.\  Janszky, C.\  Sibilia, M.\  Bertolotti, P.\  Adam and
A.\  Petak, Quantum Semiclass.\   Opt.\  {\bf 7}, 509 (1995);
J.\  Pe\v{r}ina Jr.\ and J.\  Pe\v{r}ina in {\em Progress in Optics},
Vol. 41, edited by E.\  Wolf (Elsevier, Amsterdam, 2000).
\bibitem{ou} Z.\ Y.\ Ou, L.\ J.\ Wang, X.\ Y.\ Zou, and L. Mandel,
Phys. Rev. A~{\bf 41}, R566 (1990).
\bibitem{coupl}
  M.\  L.\  Stich and M.\  Bass, {\em Laser Handbook}
  (North--Holland, Amsterdam, 1985), Chapter 4;
  A.\ Yariv and P.\ Yeh, {\em Optical Waves in Crystals}
  (J.\  Wiley, New--York, 1984);
\bibitem{hong85} C.\  K.\  Hong and L.\  Mandel, Phys.\  Rev.\  A~{\bf 31},
2409 (1985).
 \bibitem{reh-zeno} J.\  \v{R}eh\'{a}\v{c}ek, J.\  Pe\v{r}ina,
P.\  Facchi, S.\  Pascazio, and L.\  Mi\v{s}ta Jr., Phys.\  Rev.\
A~{\bf 62}, 013804 (2001).
\bibitem{others}
B.\  Misra and E.\  C.\  G.\  Sudarshan, J.\  Math.\  Phys.\  {\bf 18},
756 (1977); D.\  Home and M.\  A.\  B.\  Whitaker, Ann.\  Phys.\
{\bf 258}, 237 (1997);
P.\  Facchi and S.\  Pascazio in {\em Progress in Optics}, Vol. 41,
edited by E.\  Wolf (Elsevier, Amsterdam, 2000).
\bibitem{luis} A.\  Luis and J.\  Pe\v{r}ina, Phys.\  Rev.\  Lett.\  {\bf 76},
4340 (1996); A.\  Luis and L.\  L.\  S\'{a}nchez--Soto,
Phys.\  Rev.\  A~{\bf 57}, 781 (1998).
\bibitem{MPS} E.\  Mihokova, S.\  Pascazio and L.\ S.\  Schulman,
Phys.\  Rev.\  {\bf A56}, 25 (1997).
\bibitem{fiurasek} J.\   Fiur\'{a}\v{s}ek and J.\  Pe\v{r}ina,
Phys.\  Rev.\  A~{\bf 62}, 033808 (2000).
\bibitem{algebra} J.\ Wei and E.\ Norman, J.\ Math.\ Phys.\ {\bf 4}, 575
(1963).
\bibitem{braunstein} S.\ L.\ Braunstein, arXiv: quant-ph/9904002.
\bibitem{rehacek96} J.\ \v{R}eh\'{a}\v{c}ek and J.\ Pe\v{r}ina,
Opt.\ Commun.\ {\bf 132}, 549 (1996).
\bibitem{zeno-cont}
A.\  Peres, Am.\  J.\  Phys.\  {\bf 48}, 931 (1980); K.\  Kraus,
Found.\  Phys.\  {\bf 11}, 547 (1981); P.\  Facchi and S.\  Pascazio,
Phys.\  Rev.\  A~{\bf 62}, 023804 (2000).
\bibitem{zeno-relat}
L.\ S.\  Schulman, Phys.\  Rev.\  {\bf A57}, 1059
(1998).



\end{thebibliography}
\end{document}